\documentclass[a4paper,12pt]{article}
\usepackage {amsmath}
\usepackage {amsxtra}
\usepackage {amsbsy}
\usepackage {amscd}
\usepackage {latexsym}
\usepackage {amssymb}
\usepackage {amsfonts}
\usepackage {a4wide}
\usepackage {indentfirst}
\usepackage {bm}
\usepackage{graphicx}

\newcommand {\oks}[2]{{\raise0.7ex\hbox{${\scriptstyle #1}$}\!\mathord{\left/
{\vphantom{{1}{2}}}\right.\kern-\nulldelimiterspace}\!\lower0.7ex
\hbox{${\scriptstyle #2}$}}}

\begin{document}

\title{Neutrino quantum states in matter}
\author{Alexander Studenikin{\thanks{
Address: Department of Theoretical Physics, Moscow State
University, 119992  Moscow, Russia,
 e-mail:studenik@srd.sinp.msu.ru .}},
Alexei Ternov{\thanks{Department of Theoretical Physics, Moscow
Institute for Physics and Technology, 141700 Dolgoprudny, Russia,
e-mail:A\_Ternov@mail.ru }}}
  \date{}
  \maketitle

\begin{abstract}{We propose a modified Dirac equation
for a massive neutrino moving in the presence of the background
matter. The effects of the charged and neutral-current
interactions with the matter as well as the matter motion and
polarization are accounted for. In the particular case of the
matter with a constant density the exact solutions of this
equation are found, the neutrino energy spectrum in the matter is
also determined. On this basis the effects of the neutrino
trapping and reflection, the neutrino-antineutrino pair
annihilation and creation in a medium are studied. The quantum
theory of the spin light of neutrino in matter ($SL\nu$) is also
developed.}

\end{abstract}

\section {Introduction}

The problem of a neutrino propagation through the background
matter has attracted the permanent interest for many years. The
crucial importance of the matter effects was demonstrated in the
studies of Refs.\cite{WolPRD78,MikSmiYF85} where the resonance
amplification of the neutrino flavour oscillations in the presence
of the matter (the Mikheyev-Smirnov-Wolfenstein effect) was
discovered. The similar resonance effect in the neutrino spin
oscillations in matter was considered for the first time in
\cite{LimMarPRD88,AkhPLB88}.

It is a common knowledge that the matter effects in neutrino
oscillations have a large impact on solar-neutrino problem (see
for a recent review  \cite{BahGonPenhp0406294}); it may be of
interest in the context of neutrino oscillation processes in
supernovae and neutron stars \cite{FulMayWilSchAPJ87}. It is also
believed that the neutrino oscillations in the presence of matter
play important role in the early Universe \cite{DolYF81}.

The neutrino interaction with matter can bring about new phenomena
that do not exist in the absence of matter. In particular, we have
recently shown \cite{LobStuPLB03,LobStuPLB04} that a massive
neutrino moving in the background matter and electromagnetic
fields can produce a new type of the electromagnetic radiation. We
have named this radiation as the "spin light of neutrino"
($SL\nu$) \cite{LobStuPLB03} in order to manifest the
correspondence with the magnetic moment dependent term in the
radiation of an electron moving in a magnetic field (see
\cite{TerUFN95}). The radiation of a neutrino moving in a magnetic
field (the $SL\nu$ in a magnetic field) was studied before in
\cite{BorZhuTer88}. Recently we have also considered
\cite{DvoGriStuIJMP04} the $SL\nu$ in gravitational fields.

In \cite{LobStuPLB03,LobStuPLB04} we develop the quasi-classical
theory of the $SL\nu$ on the basis of the quasi-classical
description \cite{EgoLobStu00LobStu01DvoStuJHEP02StuPAN04} of the
neutrino spin evolution in the background matter. However, the $
SL\nu$ is a quantum process which originates from the quantum spin
flip transitions. It is important to revise the calculations of
the rate and total power of the $ SL\nu$ in matter using the
quantum theory.

The goal of this  paper is to present a reasonable step forward in
the study of a neutrino interaction in the background matter and
external fields. We derive a new quantum equation for the neutrino
wave function, in which the effects of the neutrino-matter
interaction are accounted for. This equation establishes  the
basis for the quantum treatment of a neutrino moving in the
presence of matter. In the limit of a constant matter density, we
get the exact solutions of this equation, classify them over the
neutrino spin states and determine the energy spectrum. We show
how the neutrino energy is disturbed by the presence of matter and
also find the dependence of the energy on the neutrino helicity.
From the exact expression for the neutrino energy spectrum in the
background matter it follows that, for the given neutrino
momentum, the energy of the negative-helicity neutrino in matter
exceeds the energy of the positive-helicity neutrino. In the
performed below analysis of the neutrino energy spectrum in matter
we confirm some of the results on neutrinos trapping and
spontaneous neutrino-antineutrino pair creation in a dense medium
that have been obtained previously in
\cite{ChaZiaPRD88,LoePRL90,PanPLB91-PRD92,WeiKiePRD97}.

Then with the use of the obtained neutrino wave functions in
matter we develop the quantum theory of the $ SL\nu$ and calculate
the rate and power of the spin-light radiation in matter
accounting for the emitted photos polarization. The existence of
the neutrino-spin self-polarization effect
\cite{LobStuPLB03,LobStuPLB04} in the matter is also confirmed
within the solid base of the developed quantum
approach\footnote{The neutrino-spin self-polarization effect in
the magnetic and gravitational fields were discussed in \cite
{BorZhuTer88} and \cite{DvoGriStuIJMP04}, respectively.}.

\section{Dirac equation for neutrino in matter}
To derive the quantum equation for the neutrino wave function in
the background matter we start with the effective Lagrangian that
describes the neutrino interaction with particles of the
background matter. For definiteness, we consider the case of the
electron neutrino $\nu$ propagating through moving and polarized
matter composed of only electrons (the electron gas). The
generalizations for the other flavour neutrinos and also for more
complicated matter compositions are just straightforward. Assume
that the neutrino interactions are described by the extended
standard model supplied with $SU(2)$-singlet right-handed neutrino
$\nu_{R}$. We also suppose that there is a macroscopic amount of
electrons in the scale of a neutrino de Broglie wave length.
Therefore, the interaction of a neutrino with the matter
(electrons) is coherent. In this case the averaged over the matter
electrons addition to the vacuum neutrino Lagrangian, accounting
for the charged and neutral interactions, can be written in the
form
\begin{equation}\label{Lag_f}
\Delta L_{eff}=-f^\mu \Big(\bar \nu \gamma_\mu {1+\gamma^5 \over
2} \nu \Big), \ \  f^\mu={G_F \over \sqrt2}\Big((1+4\sin^2 \theta
_W) j^\mu - \lambda ^\mu \Big),
\end{equation}
where the electrons current $j^{\mu}$ and electrons polarization
$\lambda^{\mu}$ are given by
\begin{equation}
j^\mu=(n,n{\bf v}), \label{j}
\end{equation}
and
\begin{equation} \label{lambda}
\lambda^{\mu} =\Bigg(n ({\bm \zeta} {\bf v} ), n {\bm \zeta}
\sqrt{1-v^2}+ {{n {\bf v} ({\bm \zeta} {\bf v} )} \over
{1+\sqrt{1- v^2}}}\Bigg),
\end{equation}
$\theta _{W}$ is the Weinberg angle.

 The Lagrangian (\ref{Lag_f})
accounts for the possible effect of the matter motion and
polarization. Here $n$, ${\bf v}$, and ${\bm \zeta} \ (0\leqslant
|{\bm \zeta} |^2 \leqslant 1)$ denote, respectively, the number
density of the background electrons, the speed of the reference
frame in which the mean momenta of the electrons is zero, and the
mean value of the polarization vector of the background electrons
in the above mentioned reference frame. The detailed discussion on
the determination of the electrons polarization can be found in
\cite{EgoLobStu00LobStu01DvoStuJHEP02StuPAN04}.

From the standard model Lagrangian with the extra term $\Delta
L_{eff}$ being added, we derive the following modified Dirac
equation for the neutrino moving in the background matter,
\begin{equation}\label{new}
\Big\{ i\gamma_{\mu}\big[\partial^{\mu}-\frac{1}{2}
(1+\gamma_{5})f^{\mu}\big]-m \Big\}\Psi(x)=0.
  \end{equation}
This is the most general equation of motion of a neutrino in which
the effective potential $V_{\mu}=\frac{1}{2}(1+\gamma_{5})f_{\mu}$
accounts for both the charged and neutral-current interactions
with the background matter and also for the possible effects of
the matter motion and polarization. It should be noted here that
the modified effective Dirac equations for a neutrino with various
types of interactions with the  background environment  were used
previously in \cite{ManPRD88, ChaZiaPRD88, NotRafNPB88, NiePRD89,
HaxZhaPRD91,WeiKiePRD97} for the study of the neutrino dispersion
relations and derivation of the neutrino oscillation probabilities
in matter. If we neglect the contribution of the neutral-current
interaction and possible effects of motion and polarization of the
matter then from (\ref{new}) we can get corresponding equations
for the left-handed and right-handed chiral components of the
neutrino field derived in \cite{PanPLB91-PRD92}.
\section{Neutrino wave function and energy spectrum in matter}
In the further discussion below we consider the case when no
electromagnetic field is present in the background. We also
suppose that the matter is unpolarized, $\lambda^{\mu}=0$.
Therefore, the term describing the neutrino interaction with the
matter is given by
\begin{equation}\label{f}
f^\mu=\frac{\tilde{G}_{F}}{\sqrt2}(n,n{\bf v}),
\end{equation}
where we use the notation $\tilde{G}_{F}={G}_{F}(1+4\sin^2 \theta
_W)$.

In the rest frame of the matter the equation (\ref{new}) can be
written in the Hamiltonian form,
\begin{equation}\label{H_matter}
i\frac{\partial}{\partial t}\Psi({\bf r},t)=\hat H_{matt}\Psi({\bf
r},t),
\end{equation}
where
\begin{equation}\label{H_G}
  \hat H_{matt}=\hat {\bm{\alpha}} {\bf p} + \hat {\beta}m +
  \hat V_{matt},
\end{equation}
and
\begin{equation}\label{V_matt}
\hat V_{matt}= \frac{1}{2\sqrt{2}}(1+\gamma_{5}){\tilde {G}}_{F}n,
\end{equation}
here $\bf p$ is the neutrino momentum. We use the Pauli-Dirac
representation for the Dirac matrices $\hat {\bm \alpha}$ and
$\hat {\beta}$, in which
\begin{equation}\label{a_b}
    \hat {\bm \alpha}=
\begin{pmatrix}{0}&{\hat {\bm \sigma}} \\
\hat {{\bm \sigma}}& {0}
\end{pmatrix}=\gamma_0{\bm \gamma}, \ \ \
\hat {\beta}=\begin{pmatrix}{1}&{0} \\
{0}& {-1}
\end{pmatrix}=\gamma_0,
\end{equation}
where ${\hat { \bm\sigma}}=({ \sigma}_{1},{ \sigma}_{2},{
\sigma}_{3})$ are the Pauli matrixes.

The form of the Hamiltonian (\ref{H_G}) implies that the operators
of the momentum, $\hat {\bf p}$, and longitudinal polarization,
${\hat{\bf \Sigma}} {\bf p}/p$, are the integrals of motion. So
that, in particular, we have
\begin{equation}\label{helicity}
  \frac{{\hat{\bf \Sigma}}{\bf p}}{p}
  \Psi({\bf r},t)=s\Psi({\bf r},t),
 \ \ {\hat {\bm \Sigma}}=
\begin{pmatrix}{\hat {\bm \sigma}}&{0} \\
{0}&{\hat {\bm \sigma}}
\end{pmatrix},
\end{equation}
where the values $s=\pm 1$ specify the two neutrino helicity
states, $\nu_{+}$ and  $\nu_{-}$. In the relativistic limit the
negative-helicity neutrino state is dominated by the left-handed
chiral state ($\nu_{-}\approx \nu_{L}$), whereas the
positive-helicity state is dominated by the right-handed chiral
state ($\nu_{+}\approx \nu_{R}$).

For the stationary states of the equation (\ref{new}) we get
\begin{equation}\label{stat_states}
\Psi({\bf r},t)=e^{-i(  E_{\varepsilon}t-{\bf p}{\bf r})}u({\bf
p},E_{\varepsilon}),
\end{equation}
where $u({\bf p},E_{\varepsilon})$ is independent on the
coordinates and time. Upon the condition that the equation
(\ref{new}) has a non-trivial solution, we arrive to the energy
spectrum of a neutrino moving in the background matter:
\begin{equation}\label{Energy}
  E_{\varepsilon}=\varepsilon{\sqrt{{\bf p}^{2}\Big(1-s\alpha \frac{m}{p}\Big)^{2}
  +m^2} +\alpha m} ,
\end{equation}
where we use the notation
\begin{equation}\label{alpha}
  \alpha=\frac{1}{2\sqrt{2}}{\tilde G}_{F}\frac{n}{m}.
\end{equation}
The quantity $\varepsilon=\pm 1$ splits the solutions into the two
branches that in the limit of the vanishing matter density,
$\alpha\rightarrow 0$, reproduce the positive and
negative-frequency solutions, respectively. It is also important
to note that the neutrino energy in the background matter depends
on the state of the neutrino longitudinal polarization, i.e. in
the relativistic case the left-handed and right-handed neutrinos
with equal momenta have different energies.

The procedure, similar to one used for derivation of the solution
of the Dirac equation in vacuum, can be adopted for the case of a
neutrino moving in matter. We apply this procedure to the equation
(\ref{new}) and arrive to the final form of the wave function of a
neutrino moving in the background matter:
\begin{equation}\label{wave_function}
\Psi_{\varepsilon, {\bf p},s}({\bf r},t)=\frac{e^{-i(
E_{\varepsilon}t-{\bf p}{\bf r})}}{2L^{\frac{3}{2}}}
\begin{pmatrix}{\sqrt{1+ \frac{m}{ E_{\varepsilon}-\alpha m}}}
\ \sqrt{1+s\frac{p_{3}}{p}}
\\
{s \sqrt{1+ \frac{m}{ E_{\varepsilon}-\alpha m}}} \
\sqrt{1-s\frac{p_{3}}{p}}\ \ e^{i\delta}
\\
{  s\varepsilon\sqrt{1- \frac{m}{ E_{\varepsilon}-\alpha m}}} \
\sqrt{1+s\frac{p_{3}}{p}}
\\
{\varepsilon\sqrt{1- \frac{m}{ E_{\varepsilon}-\alpha m}}} \ \
\sqrt{1-s\frac{p_{3}}{p}}\ e^{i\delta}
\end{pmatrix} ,
\end{equation}
where the energy $E_{\varepsilon}$ is given by (\ref{Energy}), and
$L$ is the normalization length. In the limit of vanishing density
of matter, when $\alpha\rightarrow 0$, the wave function
(\ref{wave_function}) transforms to the vacuum solution of the
Dirac equation.

The proposed new quantum equation (\ref{new}) for a neutrino in
the background matter, the exact solution (\ref{wave_function})
and the obtained energy spectrum (\ref{Energy}) establish a basis
for investigations of different phenomena that can appear when
neutrinos are moving in the media.

\section{Neutrino trapping and reflection in matter}

Let us now consider in some detail a neutrino energy spectrum
(\ref{Energy}) in the background matter. For the fixed magnitude
of the neutrino momentum $p$ there are the two values for the
"positive sign" ($\varepsilon =+1$) energies
\begin{equation}\label{Energy_nu}
  E^{s=+1}={\sqrt{{\bf p}^{2}\Big(1-\alpha \frac{m}{p}\Big)^{2}
  +m^2} +\alpha m}, \ \ \
 E^{s=-1}={\sqrt{{\bf p}^{2}\Big(1+\alpha \frac{m}{p}\Big)^{2}
  +m^2} +\alpha m},
\end{equation}
that determine the positive- and negative-helicity eigenstates,
respectively. The energies in Eq.(\ref{Energy_nu}) correspond to
the particle (neutrino) solutions in the background matter. The
two other values for the energy, corresponding to the negative
sign $\varepsilon =-1$, are for the antiparticle solutions. As
usual, by changing the sign of the energy, we obtain the values
\begin{equation}\label{Energy_anti_nu}
  {\tilde E}^{s=+1}={\sqrt{{\bf p}^{2}
  \Big(1-\alpha \frac{m}{p}\Big)^{2}
  +m^2} -\alpha m}, \ \ \
  {\tilde E}^{s=-1}={\sqrt{{\bf p}^{2}
  \Big(1+\alpha \frac{m}{p}\Big)^{2}
  +m^2} -\alpha m},
\end{equation}
that correspond to the positive- and negative-helicity
antineutrino states in the matter. The expressions in
Eqs.(\ref{Energy_nu}) and (\ref{Energy_anti_nu}) would reproduce
the neutrino dispersion relations of \cite{PanPLB91-PRD92} (see
also \cite{WeiKiePRD97}), if the contribution of the
neutral-current interaction to the neutrino potential were left
out.

The neutrino dispersion relations in the matter exhibits a very
fascinating feature (see also \cite{PanPLB91-PRD92, WeiKiePRD97}).
As it follows from (\ref{Energy_nu}) and (\ref{Energy_anti_nu}),
the neutrino energy may has a minimum at a non-zero momentum. It
may also happen that the neutrino group and phase velocities are
pointing  in the opposite directions.

The obtained neutrino and antineutrino energy spectra,
Eqs.(\ref{Energy_nu}) and (\ref{Energy_anti_nu}), enable us to
consider several important phenomena of a neutrino propagation in
the background medium. To illustrate the key features of the
obtained dispersion relations, we plot in Figs.1 and 2 the allowed
ranges for the neutrino and antineutrino energies for different
values of the matter density parameter $\alpha$.
\begin{figure}[h]
\begin{center}
\includegraphics[width=0.8\textwidth]{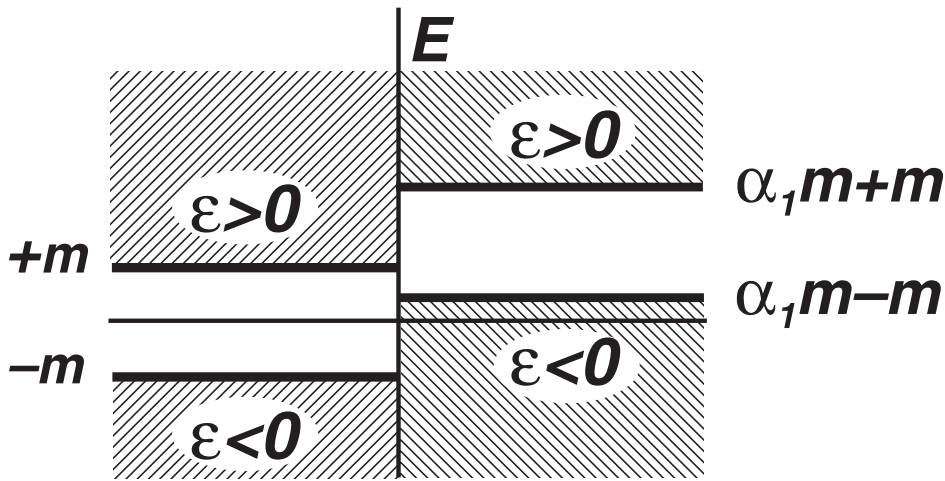}\\
\parbox{0.75\textwidth}{\caption{The ranges of the "positive"($\varepsilon=+1$) and
"negative" ($\varepsilon=-1$) sign energies in the vacuum (the
left side) and in the presence of the matter with the density
parameter $1<\alpha_{1} <2$ (the right side). The dashed and
un-dashed zones are for the allowed and forbidden values of the
energy, respectively.}\label{Fig4}}
\end{center}
\end{figure}
In the left side of Fig.1 the minimal energies for a neutrino and
antineutrino in vacuum ($\alpha=0$)  are shown by the solid lines
given by $E=\pm m$. The un-dashed gap between $E=+m$ and $E=-m$ is
the forbidden energy zone. In the right of Fig.1 the corresponding
minimal values of the neutrino and antineutrino energies in the
presence of the matter with the density parameter $1<\alpha_{1}
<2$ are shown. The forbidden energy zone is lifted up by the value
of $\alpha_{1} m$ with respect to the vacuum case. The existence
of the two interesting phenomena can be recognized from this
figure. First of all, antineutrinos with energies in the range of
$|\alpha_{1} m - m|\leq E < m$ can not escape from the medium
because this particular range of energies exactly falls on the
forbidden energy zone in the vacuum. In this case an antineutrino
has not enough energy to survive in the vacuum, therefore it is
trapped inside the medium. It should be also noted here that the
possibility for a neutrino to have the minimal energy in the
matter less than the neutrino mass and the corresponding neutrino
trapping effecct in the medium has been discussed in
\cite{WeiKiePRD97}.

The second fascinating phenomenon can appear when a neutrino is
propagating in the vacuum towards the interface between the vacuum
and the matter. We again examine the neutrino dispersion relations
illustrated by Fig.1. If the neutrino energy in the vacuum is less
than the neutrino minimal energy in the medium (this case
corresponds to the neutrino energies in the vacuum in the range
given by $m\leq E<\alpha_{1} m + m$) then the neutrino will be
reflected from the interface because the appropriate energy level
inside the medium is not accessible for the neutrino.

\section{Neutrino-antineutrino pair annihilation and creation in matter}
Let us now consider the corresponding phenomena that can appear at
the interface between the vacuum and the medium in the case when
the matter density parameter $\alpha_{2}\geq 2$ (see Fig.2).
Consider a neutrino with the energy $m<E\leq \alpha_{2} m - m$
propagating in the vacuum towards the interface with the matter.
Let us also suppose that not all of "negative sign" energy levels
in the matter are occupied and, in particular, the level with the
energy exactly equal to the energy of the neutrino which is
falling on the interface from the vacuum. This means that there is
also the antineutrino in the matter. In this case the process of
the neutrino-antineutrino annihilation can proceed at the
interface of the vacuum and the matter.

\begin{figure}[h]
\begin{center}
\includegraphics[width=0.8\textwidth]{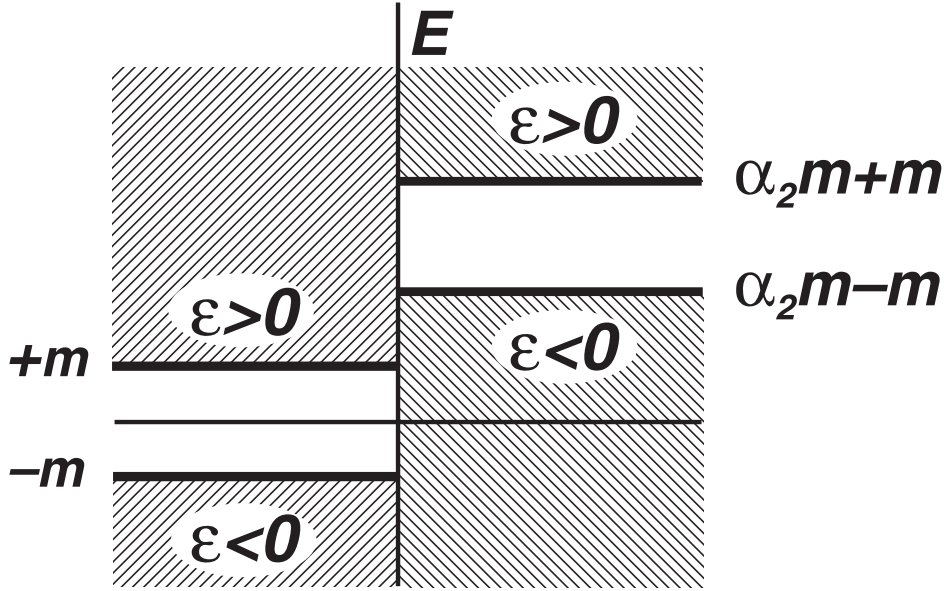}\\
\parbox{0.75\textwidth}{\caption{The ranges of the "positive"($\varepsilon=+1$) and
"negative" ($\varepsilon=-1$) sign energies in the vacuum (the
left side) and in the presence of the matter with the density
parameter $\alpha_{2}\geq 2$ (the right side). The dashed and
un-dashed zones are for the allowed and forbidden values of the
energy, respectively. }\label{Fig2}}
\end{center}
\end{figure}
From the Fig.2 that illustrates the neutrino and antineutrino
dispersion relations in the vacuum and matter with the density
parameter $\alpha_{2}\geq 2$ it can be also seen that the effect
of the spontaneous neutrino-antineutrino pair creation can appear
in the presence of the matter. Indeed, the "negative sign" energy
levels in the matter (the right-hand side of Fig.2) have their
counterparts in the "positive sign" energy levels in the vacuum
(the left-hand side of Fig.2). The neutrino-antineutrino pair
creation can be interpreted as the process of the appearance of
the particle state in the "positive sign" energy range accompanied
by the appearance of the hole state in the "negative sign" energy
sea. The phenomenon of the neutrino-antineutrino pair creation in
the presence of the matter is similar to the spontaneous
electron-positron pair creation according to Klein's paradox of
the electrodynamics (see, for instance, \cite{ItzZubQFT80}). The
possibility of the neutrino-antineutrino pair creation in the
medium was also discussed before in \cite{WeiKiePRD97,LoePRL90}.

\section {Quantum theory of spin light of neutrino in matter}
In this section we should like to use the obtained solutions
(\ref{wave_function}) of the equation (\ref{new}) for a neutrino
moving in the background matter for the study of the spin light of
neutrino ($SL\nu$) in the matter. We develop below the {\it
quantum} theory of this effect. Within the quantum approach, the
corresponding Feynman diagram of the $SL\nu$ in the matter is the
standard one-photon emission diagram with the initial and final
neutrino states described by the "broad lines" that account for
the neutrino interaction with the matter. It follows from the
usual neutrino magnetic moment interaction with the quantized
photon field, that the amplitude of the transition from the
neutrino initial state $\psi_{i}$ to the final state $\psi_{f}$,
accompanied by the emission of a photon with a momentum
$k^{\mu}=(\omega,{\bf k})$ and  polarization ${\bf e}^{*}$, can be
written in the form
\begin{equation}\label{amplitude}
  S_{f i}=-\mu \sqrt{4\pi}\int d^{4} x {\bar \psi}_{f}(x)
  ({\hat {\bm \Gamma}}{\bf e}^{*})\frac{e^{ikx}}{\sqrt{2\omega L^{3}}}
   \psi_{i}(x),
\end{equation}
where $\mu$ is the neutrino magnetic moment, $\psi_{i}$ and
$\psi_{f}$ are the corresponding exact solutions of the equation
(\ref{new}) given by (\ref{wave_function}), and
\begin{equation}\label{Gamma}
  \hat {\bm \Gamma}=i\omega\big\{\big[{\bm \Sigma} \times
  {\bm \varkappa}\big]+i\gamma^{5}{\bm \Sigma}\big\}.
\end{equation}
Here ${\bm \varkappa}={\bf k}/{\omega}$ is the unit vector
pointing the direction of the emitted photon propagation. The
integration in (\ref{amplitude}) with respect to time yields
\begin{equation}\label{amplitude}
  S_{f i}=-\mu {\sqrt {\frac {2\pi}{\omega L^{3}}}}
  2\pi\delta(E_{f}-E_{i}+\omega)
  \int d^{3} x {\bar \psi}_{f}({\bf r})({\hat {\bf \Gamma}}{\bf e}^{*})
  e^{i{\bf k}{\bf r}}
   \psi_{i}({\bf r}),
\end{equation}
where the delta-function stands for the energy conservation.
Performing the integrations over the spatial co-ordinates, we can
recover the delta-functions for the three components of the
momentum. Finally, we get the law of the energy-momentum
conservation for the considered process,
\begin{equation}\label{e_m_con}
    E_{i}=E_{f}+\omega, \ \ \
    {\bf p}_{i}={\bf p}_{f}+{\bm \varkappa}.
\end{equation}
Let us suppose that the weak interaction of the neutrino with the
electrons of the background is indeed weak. In this case, we can
expand the energy (\ref{Energy}) over ${\tilde {G}}_{F}n/p\ll 1$
and in the liner approximation get
\begin{equation}\label{Energy_2}
  E\approx E_{0}-sm\alpha \frac{p}{E_{0}}+\alpha m,
\end{equation}
where $E_0=\sqrt{p^2 +m^2}$. Then from the law of the energy
conservation (\ref{e_m_con}) we get for the energy of the emitted
photon
\begin{equation}\label{omega}
  \omega=E_{{i}_{0}}-{E}_{{f}_{0}}+\Delta,
\ \   \ \Delta=\alpha m \frac{p}{E_0}(s_{f}-s_{i}),
\end{equation}
where the indexes $i$ and $f$ label the corresponding quantities
for the neutrino in the initial and final states. From
Eq.(\ref{omega}) and the law of the momentum conservation, in the
linear approximation over $n$, we obtain
\begin{equation}\label{omega_1}
    \omega=(s_{f}-s_{i})\alpha m \frac {\beta}{1-\beta \cos
    \theta},
\end{equation}
where $\theta$ is the angle between ${\bm \varkappa}$ and the
direction of neutrino speed ${\bm \beta}$.

From the above consideration it follows that the only possibility
for the $SL\nu$ to appear is provided in the case when the
neutrino initial and final states are characterized by $s_{i}=-1$
and $s_{f}=+1$, respectively. Thus,  on the basis of the quantum
treatment of the $SL\nu$ in the matter we conclude, that in this
process the left-handed neutrino is converted to the right-handed
neutrino (see also \cite{LobStuPLB03}) and the emitted photon
energy is given by
\begin{equation}\label{omega_12}
    \omega=\frac{1}{\sqrt{2}}{\tilde G}_{F}n \frac {\beta}{1-\beta \cos
    \theta}.
\end{equation}
Note that the photon energy depends on the angle $\theta$ and also
on the value of the neutrino speed $\beta$. In the case of
$\beta\approx 1$ and $\theta\approx 0$ we confirm the estimation
for  the emitted photon energy given in \cite{LobStuPLB03}.

If further we consider the case of the neutrino moving along the
OZ-axes, we can rewrite the solution (\ref{wave_function}) for the
neutrino states with $s=-1$ and $s=+1$ in the following forms
\begin{equation}\label{wave_function_min}
\Psi_{{\bf p},s=-1}({\bf r},t)=\frac{e^{-i(Et-{\bf p}{\bf
r})}}{\sqrt{2}L^{\frac{3}{2}}}
\begin{pmatrix}{0}
\\
{- \sqrt{1+ \frac{m}{E-\alpha m}}} \
\\
{ 0}
\\
{\sqrt{1- \frac{m}{E-\alpha m}}}
\end{pmatrix} ,
\end{equation}
and
\begin{equation}\label{wave_function_min}
\Psi_{{\bf p},s=+1}({\bf r},t)=\frac{e^{-i(Et-{\bf p}{\bf
r})}}{\sqrt{2}L^{\frac{3}{2}}}
\begin{pmatrix}{ \sqrt{1+ \frac{m}{E-\alpha m}}}
\\
{0} \
\\
{ \sqrt{1- \frac{m}{E-\alpha m}}}
\\
{0}
\end{pmatrix}.
\end{equation}
We now put these wave functions into Eq.(\ref{amplitude}) and
calculate the spin light transition rate in the linear
approximation of the expansion over the parameter ${\tilde
{G}}_{F}n/p$. Finally, for the rate we get
\begin{equation}\label{rate}
  \Gamma_{SL}=\frac{1}{2\sqrt{2}}{\tilde G}_{F}^{3}\mu^{2}n^{3}
  \beta^{3}
\int
  \limits_{}^{}\frac{S\sin \theta}{(1-\beta \cos \theta)^{4}}
   d\theta.
\end{equation}
where
\begin{equation}\label{S}
S=(\cos \theta - \beta)^{2}+
  (1-\beta \cos \theta)^{2}.
\end{equation}
Performing the integrations in Eq.(\ref{rate}) over the angle
$\theta$, we obtain for the rate
\begin{equation}\label{rate_1}
  \Gamma_{SL}=\frac{2\sqrt{2}}{3}\mu^{2}{{\tilde G}_{F}}^{3}
  n^{3}\beta^{3}\gamma^{2}.
\end{equation}
This result exceeds the value of the neutrino spin light rate
derived in \cite{LobStuPLB03} by a factor of two because here the
neutrinos in the initial state  are totally left-hand polarized,
whereas the case of the unpolarized neutrinos in the initial state
was consider in \cite{LobStuPLB03}.

The corresponding expression for the radiation power is
\begin{equation}\label{power}
  I_{SL}=\frac{1}{4}\mu^{2}{\tilde G}_{F}^{4}n^{4}\beta^{4}
\int\limits_{}^{}\frac{S\sin \theta}{(1-\beta \cos \theta)^{5}}
   d\theta.
\end{equation}
Performing the integration, we get for the total radiation power
\begin{equation}\label{power_1}
  I_{SL}=\frac{2}{3}\mu^{2}{{\tilde G}_{F}}^{4}
  n^{4}\beta^{4}\gamma^{4}.
\end{equation}

In the performed above quantum treatment of the $SL\nu$ in the
background matter we confirm the main properties
\cite{LobStuPLB03, LobStuPLB04, DvoGriStuIJMP04} of this
radiation. In particular, as it follows from (\ref{power}), the
$SL\nu$ is strongly beamed along the propagation of the
relativistic neutrino. The total power of the $SL\nu$ in the
matter is increasing with the increase of the background matter
density and the neutrino $\gamma$ factor, $I_{SL}\sim n^{4}\gamma
^{4}$. From the obtained within the quantum treatment expression
(\ref{omega_12})  one can get an estimation for the emitted photon
energy
\begin{equation} \omega =2.37\times
10^{-7}\left( \frac{n}{10^{30}cm^{-3}}\right) \left(
\frac{E}{m_{\nu }}\right) ^{2}eV.
\end{equation}
It follows that for the matter densities and neutrino energies,
appropriate for the neutron star environments, the range of the
radiated photons energies may span up to the gamma-rays.

Using the neutrino transition amplitude  in the matter,
Eq.(\ref{amplitude}), it is also possible \cite{StuTerQUARKS04} to
derive the $SL\nu$ rate and power with the emitted photons
polarizations being accounted for. Information on the photons
polarization may be important for the experimental observation of
the $SL\nu$ from different astrophysical and cosmology objects and
media.

Finally, the developed above quantum theory of the neutrino motion
in the background matter also reveals the nature of the $SL\nu$.
In particular, the application of the quantum theory to this
phenomenon enables us to demonstrate that the $SL\nu$ appears due
to the two subdivided phenomena: (i) the shift of the neutrino
energy levels in the presence of the background matter, which is
different for the two opposite neutrino helicity states, (ii) the
radiation of the $SL\nu$ photon in the process of the neutrino
transition from the "exited" negative-helicity state to the
low-lying positive-helicity neutrino state in matter. Therefore,
as it has been discussed in \cite{LobStuPLB03, LobStuPLB04}, the
relativistic neutrino beam composed of the active neutrinos
$\nu_{L}$ can be converted to the sterile neutrinos $\nu_{R}$.

\section{Summary}
The quantum approach to description of a neutrino moving in the
background matter based on the modified Dirac equation has been
proposed. The exact solutions of this equation and the neutrino
energy spectrum have been derived in the case of a matter with
constant density. The neutrino trapping and reflection, and also
the neutrino-antineutrino pair annihilation and creation in a
medium have been studied. The quantum theory of the spin light of
a neutrino in the matter has been also developed. The quantum
method for including a matter background would have a large impact
on the studies of different processes with neutrinos propagating
in the astrophysical and cosmology media.

\section{Acknowledgement}
The authors thanks Alexander Grigoriev and Andrey Lobanov for
discussion on some of the issues of the present paper.

\end{document}